\documentclass[%
 aip,
 amsmath,amssymb,
reprint,%
]{revtex4-1}

\usepackage{graphicx}
\usepackage{dcolumn}
\usepackage{bm}

\usepackage[utf8]{inputenc}
\usepackage[T1]{fontenc}
\usepackage{amsmath}
\usepackage{mathptmx}
\usepackage{etoolbox}
\usepackage{multirow}
\usepackage[hidelinks]{hyperref}
\usepackage{subcaption}
\usepackage{orcidlink}

\usepackage[final]{changes}

\makeatletter
\def\@email#1#2{%
 \endgroup
 \patchcmd{\titleblock@produce}
  {\frontmatter@RRAPformat}
  {\frontmatter@RRAPformat{\produce@RRAP{*#1\href{mailto:#2}{#2}}}\frontmatter@RRAPformat}
  {}{}
}%
\makeatother

\usepackage{comment} 

\begin{document}

\onecolumngrid
\noindent The following article has been accepted by The Journal of Chemical Physics. After it is published, it will be found at \hyperlink{https://doi.org/10.1063/5.0311040}{https://doi.org/10.1063/5.0311040}.

\title{Relativistic quintuple-zeta basis sets for the p block}
\author{Marten L. Reitsma\orcidlink{0000-0002-8255-7480}}
\affiliation{Van Swinderen Institute for Particle Physics and Gravity, University of Groningen, Nijenborgh 4, 9747 AG Groningen, The Netherlands}
\author{Eifion H. Prinsen\orcidlink{0009-0008-9660-7152}}
\affiliation{Van Swinderen Institute for Particle Physics and Gravity, University of Groningen, Nijenborgh 4, 9747 AG Groningen, The Netherlands}
\author{Johan D. Polet\orcidlink{0009-0000-2751-5635}}
\affiliation{Van Swinderen Institute for Particle Physics and Gravity, University of Groningen, Nijenborgh 4, 9747 AG Groningen, The Netherlands}
\author{Anastasia Borschevsky\orcidlink{0000-0002-6558-1921}}
\affiliation{Van Swinderen Institute for Particle Physics and Gravity, University of Groningen, Nijenborgh 4, 9747 AG Groningen, The Netherlands}
\author{Kenneth G. Dyall\orcidlink{0000-0002-5682-3132}}
\affiliation{Dirac Solutions, 10527 NW Lost Park Drive, Portland, Oregon 97229, USA}
\email{diracsolutions@gmail.com}

\def\Eh{$E_{\rm h}$}
\def\uEh{$\mu E_{\rm h}$}
\def\ph{\phantom{0}}
\def\phm{\phantom{-}}

\begin{abstract}
Relativistic quintuple-zeta basis sets for the p elements are presented. The basis sets for the occupied spinors were optimized at the Dirac-Coulomb self-consistent field (SCF) level on the ground configurations. Valence and core correlating functions were optimized in multireference SDCI calculations on the ground configuration. Diffuse functions optimized on the anion (or derived from neighboring elements for group 18) are also provided. Basic atomic and molecular properties were used to test the newly developed basis sets.
A smooth convergence to the basis set limit is observed with increased basis set quality from the previously available double-zeta, triple-zeta, and quadruple-zeta basis sets to the newly developed quintuple-zeta basis sets for the calculated molecular bond lengths and dissociation energies and for atomic ionization potentials and electron affinities. Use of these basis sets in combination with state-of-the-art approaches for treatment of relativity and correlation will allow significantly increased accuracy in calculations on the heavy elements and their compounds. The basis sets are available from zenodo.org.

\end{abstract}

\maketitle

\section{Introduction}

Spectroscopic and precision experiments on heavy, unstable, and artificial atoms and molecules containing such elements provide unique opportunities to investigate the electronic structure in the lower part of the periodic table, probe the nuclear structure of heavy nuclei and unstable isotopes, test the Standard Model of particle physics, and search for the tiny signatures of physics beyond the Standard Model.~\cite{CamMooPea16, BloLaaRae21, SafBudDeM18,Arrowsmith-Kron_2024} Such experiments are also inherently challenging, due to the minute quantities and short lifetimes of the investigated elements or to the unprecedented sensitivity required in order to detect the vanishingly small signatures of new physics. Success of these ambitious experiments thus requires dedicated facilities and specially developed experimental techniques.

However, another crucial factor in these challenging experiments is the availability of strong and reliable theoretical support. Electronic structure theory can provide invaluable contributions in all stages of the experiment, from conception and planning to the interpretation of the results.  Dealing with heavy many-electron systems requires state-of-the-art treatment of both relativistic effects and electron correlation. To be useful for planning and interpreting experiments, the theoretical predictions should also have quantifiable and reliable uncertainties. Continuous development and improvement of computational methods, such as the relativistic coupled cluster approach, now allows us to reach meV level accuracy for atomic and molecular transition energies and ionization potentials, and an accuracy of a few percent for a variety of properties (see, e.g., Refs.~\onlinecite{Pasteka:17, LeiKarGuo20, Skr21, HaaEliIli20,PasEliRei26}). Such accuracy can be reached by employing state-of-the art treatment of correlation effects and going beyond the four-component framework and incorporating higher-order relativistic corrections (Breit contribution and Lamb shift). 

A balanced high-accuracy calculation also requires the use of high-quality, converged basis sets. Currently, the accuracy of the calculated values is often limited by the quality of the available basis sets.~\cite{LeiKarGuo20,GusRicRei20, KanYanBis20,GuoPasEli21,GuoBorEli22} 
This limitation serves as a motivation for the development of higher-quality basis sets that will allow us to push the accuracy of atomic and molecular calculations even further. For heavy elements, these basis sets must be capable of describing atoms in which relativistic effects, both spin-free and spin-dependent, have a significant impact on the wave function. 

Basis set calculations that are saturated with respect to both the SCF orbital representation and electron correlation are in many cases prohibitively expensive. To approach the accuracy of such calculations, it is common to perform calculations with a series of basis sets whose size and quality increase along the series, and extrapolate the results to the complete basis set (CBS) limit. The correlation-consistent basis sets~\cite{cc1,cc2,cc3,cc4} provide a systematic means for doing precisely this task. These basis sets have been extended into the relativistic domain with spin-free relativistic Hamiltonians, though the coverage is not complete.~\cite{deJong,Peterson3d,Peterson4d,Peterson5d,Peterson-slight,Peterson5p6p,PetPuzz,PetersonLn,PetersonAc1,PetersonAc2,Peterson-sblock} For heavy-element calculations, the Dyall basis sets~\cite{dyall_pdz,dyall_ptz,dyall_5d,dyall_pqz,dyall_4d,dyall_5f,dyall_s,dyall_5drev,dyall_4f,dyall_6d,dyall_7p,dyall_core_corr,dyall_light} are commonly used, as they cover all elements up to Z=118 at the double-zeta (2z), triple-zeta (3z), and quadruple-zeta (4z) levels. Extrapolations from the 3z and 4z basis set results traditionally provide an estimate of the CBS limit; however, the accuracy of the extrapolations based on these two sets was often found to be the limiting factor for comparison with experiment.~\cite{LeiKarGuo20,GusRicRei20, KanYanBis20,GuoPasEli21,GuoBorEli22} In order to improve the accuracy, basis sets of quintuple-zeta (5z) quality are thus needed. 

As a first step in filling this need, relativistic 5z basis sets for the s-block elements have recently been published.~\cite{dyall_s_5z} This paper reports the development and testing of relativistic 5z basis sets for the p-block elements. In addition to SCF sets for the occupied orbitals, these include correlating functions for the valence and core orbitals and diffuse functions for negative ion character and dipole polarization. These basis sets complement the previously developed 2z, 3z, and 4z basis sets for these elements,~\cite{dyall_pdz,dyall_ptz,dyall_pqz,dyall_7p,dyall_core_corr,dyall_light} and thus can be combined in CBS extrapolation schemes. 
The performance of the newly developed basis sets and the quality of the extrapolation to CBS limit is illustrated in calculations of basic atomic and molecular properties.

\section{Methods}

The methods used in the development of this family of basis sets have been described previously.~\cite{dyall_pdz,dyall_ptz,DyFaeBas,Sethetal,DyallCCC} The specific approaches used for the p-block basis sets~\cite{dyall_pqz} are summarized below, along with some new strategies. The basis set extensions of the General Relativistic Atomic Structure Package GRASP~\cite{DyFaeBas} and the RAMCI program~\cite{Sethetal} were used for the basis set optimizations of the SCF and correlating sets respectively.

\subsection{SCF sets}

The SCF basis sets are optimized within Dirac-Hartree-Fock (DHF) calculations using the Dirac-Coulomb Hamiltonian with the standard Gaussian nuclear charge distribution,~\cite{VissDy} to minimize the average energy of the ground configuration. As for the previously developed basis sets, $\ell$-optimization was employed, i.e. a separate set of exponents is optimized for each $\ell$ value. 

As the energy gains for increasing the basis set size from 4z to 5z are relatively small, the criterion of energy balance used in the smaller basis sets is no longer applicable. Instead, basis sets of a series of sizes were optimized, and the final basis sets were chosen as the sets for which there were five functions contributing significantly to the outermost maxima of the s and p orbitals, and at least five for the d and f orbitals. Where possible, sets of the same size were chosen for an entire row.

To increase the size of the basis set from 4z to 5z, the 4z basis sets were taken as a starting point for group 18. After optimization, the group 18 exponents are then scaled down and used as the starting set for group 17. This procedure of using the optimized exponents of group $m$ as the starting set of group $m-1$ is continued down to group 13. As a general rule, increasing the zeta level involves adding one exponent for each outermost orbital maximum except for the first (i.e., $n=\ell+1$), and then adding a number of exponents for the first, which may be any number up to the total number added to all the other maxima. 
The exact number added depends on two factors: the extent of the nucleus and the extent of the orbitals. For increasing Z, the extent of the innermost orbitals decreases and the nuclear radius increases. \replaced{The decreasing amount of space between the nucleus and the first radial maximum means that there is less space for Gaussians to contribute in this region, and consequently fewer Gaussians are needed to increase the zeta level for high Z than for low Z.}{, and there is more overlap of the Gaussians that describe these orbitals with the nuclear charge distribution The increasing overlap reduces the potential energy relative to the kinetic energy, and the minimum in the energy with respect to the exponents shifts to smaller exponent values. As a consequence, fewer exponents can be added to the inner orbitals as $Z$ increases.}

A new strategy, described in detail in Ref.~\onlinecite{dyall_s_5z}, was used for the addition of exponents, in which the exponents in a given range are replaced with an even-tempered set, increasing the number of exponents and expanding the range covered by the previous exponents.
Due to the large size and high exponent density of the 5z basis sets, the numerical accuracy is lowered due to a larger degree of linear dependence compared to the smaller basis sets. 
The lower accuracy limits the convergence of the SCF calculations and thereby limits the tolerance to which the exponent optimization can be converged. For the heaviest elements, optimizing all the exponents simultaneously often resulted in little or no improvement, so the optimization was done on groups of adjacent exponents, cycling through the entire set by groups. 

In line with the strategy used for the smaller basis sets, the d sets for the 4p, 5p, 6p, and 7p blocks were optimized with the outermost exponents taken from the valence correlating set and held fixed. The number of correlating exponents taken varied for each block, due to the relativistic expansion of the $(n-1)$d orbital. For the 4p, 5p, and 6p blocks, the smallest three correlating exponents were taken. \deleted{This also fixed a problem with coalescence of the two largest correlating exponents during the optimization of the four d functions, as the largest correlating exponent is replaced by a suitable SCF exponent.} For the 7p block, the smallest two correlating exponents were taken. This was the best compromise between SCF and correlated optimization, due to the near-complete overlap of the correlating exponent range and 6d SCF exponent range. Similarly, the f set for the 6p and 7p elements was optimized with the three valence correlating f exponents added and held fixed.

\subsection{Correlating sets}

Correlating functions were determined in the style of the correlation-consistent basis sets~\cite{cc1}, where for an increase of one in the zeta level, the maximum angular momentum represented in the correlating set is increased by one and the number of functions for each $\ell$ is increased by one.

Valence correlation functions were optimized in MR-SDCI calculations on the average energy of the valence s$^2$p$^n$ states, with a 4s4p4d3f2g1h correlating space. The d, f, g, and h exponents were optimized, the s and p exponents were chosen from the SCF set and held fixed. The s correlating exponents were chosen to be the first, second, fourth, and fifth outermost exponents from the SCF set. The p exponents for the 4p, 5p and 6p blocks were similarly taken as the first, second, fourth, and fifth outermost exponents. For the 7p block, the second, third, fifth, and sixth exponents were taken: due to the large spin-orbit splitting, the exponent sets for the outer antinode of the p$_{1/2}$ and p$_{3/2}$ spinors overlap but do not coincide. As explained above, the larger d exponents were replaced with exponents from the SCF set, due to the overlap of the SCF and correlating sets. 

Correlating functions for each inner shell were optimized in MR-SDCI calculations with an active space taken from orbitals in the target shell. The angular momentum resulting from the coupling of the vacancies in these shells to the correlating functions was constrained to a zero value.  This choice reduces the length of the CI expansion used in the exponent optimization and eliminates coupling with the valence shell. The active space was taken to include all orbitals in the shell, except when the shell included an occupied f orbital. In this case, the active space included only the f orbital. The correlating space depends on the highest angular momentum  $\ell_\mathrm{max}$ of the orbitals in the shell: for $\ell_\mathrm{max}=0$, the correlating space is 4s4p3d2f1g; for $\ell_\mathrm{max}=1$, the correlating space is 4s4p4d3f2g1h; for $\ell_\mathrm{max}=2$, the correlating space is 4s4p4d4f3g2h1i; and for $\ell_\mathrm{max}=3$, the correlating space is 4g3h2i1k. The last choice of correlating space was made to reduce the length of the CI expansion by eliminating the occupied symmetries\added{, thus making the calculations more tractable}. For the other active spaces, the correlating functions for the occupied symmetries were chosen as for the valence, with the first, second, fourth, and fifth exponents for an antinode added to the correlating set and held fixed during the optimization.

In a number of cases, the range of exponents in the unoccupied symmetries for a shell overlapped with the range for an adjacent shell. Here, a replacement was made in one set from the other, and the replacements were held fixed and the remainder of the set reoptimized, as described below. 

For the 3p block, the d function range for the $n=3$ correlating set overlapped with those for the $n=2$ shell. The largest d exponent from the $n=3$ correlating set was used as the smallest d exponent for the $n=2$ correlating set.

For Se, Br and Kr in the 4p block, the f functions for the 3d overlapped with those for the $n=2$ shell, therefore the largest f exponent from 3d was used as the smallest f exponent for the $n=2$ correlating set. There was no overlap for the remaining 4p elements, so those have an additional independent f exponent for the $n=2$ correlating set.

For the 5p block, the f function range for the 3d overlapped the ranges for both the $n=2$ and $n=4$ shells. The largest f exponent from the 3d correlating set was used as the smallest f exponent in the $n=2$ correlating set, and the smallest f exponent from the 3d correlating set was used as the largest f exponent in the $n=4$ correlating set.

For the 6p and 7p blocks, the g function range for the 4f overlapped the ranges for both the $n=3$ and $n=5$ shells. The largest g exponent from the 4f correlating set was used as the smallest g exponent in the $n=3$ correlating set, which was reoptimized with the replacement held fixed. This replacement also addresses an issue with the coalescence of two exponents during the full optimization of the $n=3$ correlating set: with a fixed exponent of a different value, the second of the two does not collapse onto the fixed one. The smallest g exponent from the 4f correlating set was used as the largest g exponent in the $n=5$ correlating set for the 6p block. The smallest two g exponents from the 4f correlating set were used as the largest two g exponents in the 5f correlating set for the 7p block. The h sets also overlap for the $n=3$ and $n=4$ shells; here the largest h function for $n=4$ was taken as the smallest h function for $n=3$ and held fixed.

For the 6p and 7p blocks, the range of f exponents for 1s correlation overlaps but is not entirely included in the range of the SCF f set. The smaller of the two f exponents is close to the largest exponent in the f SCF set, so the smaller f correlating exponent was replaced by the largest SCF exponent, and the larger correlating exponent was reoptimized.

\subsection{Diffuse functions}

One diffuse $s$ function and one diffuse $p$ function were determined in SCF calculations on the negative ion for all the elements except those in group 18. The SCF calculations were performed on the s$^2$p$^{n+1}$ configurations. Diffuse d, f, g, and h functions (one in each symmetry) were optimized for the negative ion in MR-SDCI calculations with a 5s5p5d4f3g2h correlating space consisting of the valence 4s4p4d3f2g1h set supplemented by a 1s1p1d1f1g1h diffuse set, with the diffuse s and p functions taken from the SCF negative ion optimizations and held fixed. 

The exponents of the diffuse functions for the group 18 elements were determined for each symmetry by taking the ratio of the diffuse exponent to the smallest exponent for the neutral atom from the group 17 element, and applying it to the smallest exponent of the group 18 element.

\subsection{Outer core polarization functions}

Outer core polarization is covered adequately by the valence correlating set, with the exception of the highest symmetry. An i function for correlated polarization of the outer core $(n-1)$d shell for the 4p through 7p blocks was generated. The exponent was determined by taking the ratio of the polarizing f exponent to the correlating f exponent for the $(n-1)$d shell in the double-zeta basis set, and applying the ratio to the exponent of the correlating i function for the $(n-1)$d shell.

\section{Basis set results}

A summary of the SCF basis set definitions is given in Table~\ref{basis-sizes}. This summary includes the valence correlating functions for the occupied symmetries (d and f), as these functions were used in the determination of the SCF exponents and contribute significantly to the SCF energy. The total SCF and valence MR-SDCI configuration average energies for the basis sets are given in Table~\ref{basis-energies}. Both sets of energies are degeneracy-weighted averages over the states of the s$^2$p$^n$ configuration. The active space for the CI contains the $n$s and $n$p orbitals, and the virtual space consists of the $4s4p4d3f2g1h$ positive-energy virtuals resulting from diagonalizing the DHF matrix in the space of the occupied virtuals and the valence correlating $4s4p4d3f2g1h$ primitive set.  These energies are provided for the purpose of validating the implementation of the basis sets in an installation of four-component relativistic molecular software.

\begin{table}[ht!]
 \begin{center}         
 \caption{SCF basis set sizes for the p-block elements.}\label{basis-sizes}
 \centering                                                         
 \begin{tabular}{lllll}
 \hline
 \noalign{\vspace{4pt}}
 Block & Elements & Basis \\
 \noalign{\vspace{4pt}}
 \hline
 \noalign{\vspace{4pt}}
 2p & B & 23s 12p \\
  & C & 22s 11p \\
  & N, O & 21s 11p \\
  & F, Ne & 20s 11p \\
 3p & Al -- Ar & 28s 18p\\
 4p & Ga -- Kr & 35s 26p 18d \\ 
 5p & In -- Xe & 38s 32p 23d \\
 6p & Tl -- Rn & 38s 38p 24d 16f \\
 7p & Nh -- Og & 39s 42p 30d 19f \\
 \noalign{\vspace{4pt}}
 \hline
 \end{tabular}
 \end{center}
\end{table}

\begin{table*}[ht!]
 \begin{center}
 \caption{SCF and valence MR-SDCI configuration average energies in hartrees for the p-block elements, calculated with the 5z basis sets.}\label{basis-energies}
 \begin{tabular}{lrrclrr}
 \hline
 \noalign{\vspace{4pt}}
 Element & SCF\hspace{2pc} & Valence MR-SDCI &  & Element & SCF\hspace{2pc} & Valence MR-SDCI \\
 \noalign{\vspace{4pt}}
 \hline
 \noalign{\vspace{4pt}}
 $^{11}$B  &  $-24.5365541$ &  $-24.5801943$ && $^{115}$In & $-5880.4315876$ & $-5880.4616558$ \\
 $^{12}$C  &  $-37.6760400$ &  $-37.7603077$ && $^{120}$Sn & $-6176.1280955$ & $-6176.1852919$ \\
 $^{14}$N  &  $-54.3277202$ &  $-54.4585879$ && $^{121}$Sb & $-6480.5186347$ & $-6480.6063322$ \\
 $^{16}$O  &  $-74.8249828$ &  $-75.0070222$ && $^{130}$Te & $-6793.6989756$ & $-6793.8184117$ \\
 $^{19}$F  &  $-99.5016090$ &  $-99.7378575$ && $^{127}$I  & $-7115.7941858$ & $-7115.9452235$ \\
 $^{20}$Ne & $-128.6919203$ & $-128.9842164$ && $^{132}$Xe & $-7446.8954524$ & $-7447.0765508$ \\
 \noalign{\vspace{4pt}}
 $^{27}$Al & $-242.3307487$ & $-242.3699256$ && $^{205}$Tl & $-20274.8511117$ & $-20274.8768441$\\
 $^{28}$Si & $-289.4613370$ & $-289.5366351$ && $^{208}$Pb & $-20913.7148598$ & $-20913.7640201$\\
 $^{31}$P  & $-341.4946677$ & $-341.6105013$ && $^{209}$Bi & $-21565.7066737$ & $-21565.7829328$\\
 $^{32}$S  & $-398.5979290$ & $-398.7564557$ && $^{209}$Po & $-22230.9977471$ & $-22231.1026349$\\
 $^{35}$Cl & $-460.9383788$ & $-461.1402342$ && $^{210}$At & $-22909.8083626$ & $-22909.9418088$\\
 $^{40}$Ar & $-528.6837610$ & $-528.9285991$ && $^{222}$Rn & $-23602.1051032$ & $-23602.2654763$\\
 \noalign{\vspace{4pt}}
 $^{69}$Ga & $-1942.5637613$ & $-1942.5978663$ && $^{287}$Nh & $-48511.8338983$ & $-48511.8532976$ \\
 $^{74}$Ge & $-2097.4703590$ & $-2097.5352542$ && $^{289}$Fl & $-49718.4624648$ & $-49718.5430812$\\
 $^{75}$As & $-2259.4419110$ & \replaced{$-2259.5414557$}{$-2258.5368094$} && $^{291}$Mc & $-50950.6995479$ & $-50950.8066118$\\
 $^{80}$Se & $-2428.5882731$ & $-2428.7241826$ && $^{293}$Lv & $-52209.3657095$ & $-52209.4786571$\\
 $^{79}$Br & $-2605.0234845$ & $-2605.1958111$ && $^{296}$Ts & $-53495.0179294$ & $-53495.1293394$\\
 $^{84}$Kr & $-2788.8606229$ & $-2789.0683250$ && $^{300}$Og & $-54808.4894988$ & $-54808.6229557$\\
 \noalign{\vspace{4pt}}
 \hline
 \end{tabular}
 \end{center}
\end{table*}

For the heavy elements, the shells between the innermost shell and the outer core shells can be quite compact, and thus the ratio between adjacent exponents has to be smaller than for the valence or outer core shells, or for the innermost core shell. For an example of the exponent ratios, see Fig.~1 in the review by Dyall.~\cite{DyallCCC} The small ratios can lead to linear dependence problems, and prevent the optimization of five unique exponents for the outermost maximum of each core orbital. Moreover, there is sufficient overlap between the outermost maxima for adjacent shells that the smallest exponent for the maximum of one shell is suitable for the largest exponent for the next highest shell; in other words, the exponent is shared between two shells. The linear dependence issue is largest for the s shell, as the magnitude of the overlap integral S for two Gaussians with exponents $\zeta_i$ and $\zeta_j$, defined by
\begin{equation}
    S_{ij;l} = \left( \frac{2\sqrt{\zeta_i\zeta_j}}{\zeta_i+\zeta_j} \right)^{\ell+3/2}
\end{equation}
decreases with increasing $\ell$. The larger the overlap, the greater the linear dependence, hence s Gaussians have the greatest linear dependence issues.

For the 2p elements, it was impossible to choose SCF sets of the same size for the entire block and maintain the same quality of description (zeta level) of the outer 2s antinode. This is largely because the inner s exponents increase much less across the block from B to Ne than the outer exponents: a factor of 1.6 for the inner exponents and 4.6 for the outer exponents. The same is not true for the p exponents, where the increase is about a factor of 4 for both the inner and outer exponents. The difference in behavior can be attributed to a saturation of the s set near the nucleus. The large changes in the smallest exponent are essentially due to the doubling of the nuclear charge from B to Ne. There is not such a large change from Al to Ar for the 3p elements, and less of a discrepancy between the changes in the inner and outer s exponents. Hence it was possible to choose sets of the same size for all elements in the 3p block, and for the same reason, also for the 4p through 7p blocks.

For a 5z basis set, we would expect the number of functions in a given symmetry to increase by at least five, going from one block to the next, due to the additional shell. In addition, the number of functions to describe the innermost orbital ($n=\ell+1$) also changes. Thus from the 2p block to the 3p block, we see an increase of eight s functions and seven p functions, of which five contribute to the new $n=3$ shell, and the remainder go into the innermost shell, due to the increased nuclear charge. From the 3p shell to the 4p shell, seven s and eight p functions are added. The next addition from 4p to 5p is three s functions and six p functions. The reduction in the number of s functions added is due to the saturation of the s set near the nucleus as the nuclear size increases, and to sharing of exponents between shells. From 5p to 6p, no s functions are added, but six p functions are added. There is more sharing of exponents in the s set, and the number of functions for the 1s decreases. By contrast, the p set is not yet saturated near the nucleus. From 6p to 7p, one s function is added and four p functions. Again the number of functions for the 1s decreases, and the p set is starting to become saturated near the nucleus. The p set is now larger than the s set. There are two factors contributing to this. First, the p set must describe the small component of the 2p$_{1/2}$ spinor, which is s-like and contributes a substantial percentage of the orbital density. It therefore must cover a similar range of exponents near the nucleus as the s set. In fact the largest exponent is larger than for the s set---a feature already seen for the quadruple-zeta basis sets (see Fig.~3 in the review by Dyall~\cite{DyallCCC}). Second, the contraction of the 7s and the expansion of the 7p$_{3/2}$ due to spin-orbit splitting mean that the outermost exponents of the p set are smaller than those of the s set. This accounts for one extra function on each end of the p set, compared to the s set. The final extra function compared to the s set contributes to the 2p orbital. 

The spin-orbit splitting is sufficiently large that the exponents for the 7p$_{1/2}$ cover the same range as the 7s, and there is a shift in the nodal pattern between the 7p$_{1/2}$ and the 7p$_{3/2}$ by one exponent. This was also observed for the quadruple zeta basis set. It suggests that $\ell$-optimization is not the best strategy for these elements, and it may ultimately be more efficient to use $j$-optimization.~\cite{DyFaeBas}

\section{Applications}
To investigate the performance of the newly-developed basis sets, calculations were carried out on several atomic and molecular systems. In order to show the consistency of the new basis sets with the previous ones, we compared calculations with the 5z basis sets to those obtained with the 2z, 3z and 4z basis sets. Additionally, the extrapolation to the complete basis set (CBS) limit from different basis set qualities is demonstrated. The CBS(4z) limit is determined from calculations using the 3z and 4z basis sets, while CBS(5z) also includes the 5z basis set results. The 2z basis sets were not included in the fits as the quality of the results is insufficient and would distort the extrapolation. However, the 2z basis sets are shown in the plots presented below. 

The calculations were performed at the Dirac-Coulomb level using the SCF, CCSD and CCSD(T) methods, using either a valence basis set (v) that includes only functions correlating the outer $n$ shell, or a core-valence (cv) set that also includes correlating functions for the ($n-1$) core shell. Basis sets that start with the letter ``a'' include the diffuse functions. For example, the basis set acv5z is the 5-zeta basis set that includes diffuse functions as well as correlating functions for the $n$ and ($n-1$) shells. 
The SCF energies are not extrapolated, because they are converged sufficiently for the larger basis sets and the correlation energy is the dominant contribution to the CBS limit. This means that we set $E^{\text{SCF}}_{\text{CBS(4z)}}$~=~$E^{\text{SCF}}_{\text{4z}}$ and $E^{\text{SCF}}_{\text{CBS(5z)}}$~=~$E^{\text{SCF}}_{\text{5z}}$. The correlation energies are extrapolated using the scheme proposed by Martin,~\cite{Martin1996}
\begin{equation}
E_N^{\text{corr}} = E_{\text{CBS}}^{\text{corr}} + \frac{A}{N^3},
\end{equation}
where $N$ is the basis set cardinality. The CBS(4z) or CBS(5z) limits were obtained from a fit of the energies obtained at the 3z and 4z, or the 3z, 4z, and 5z basis set cardinalities, respectively. The CBS limit for the properties were determined from the extrapolated energies (for the IPs and EAs) or the extrapolated potential energy curves (in case of the bond lengths and dissociation energies).

Results of the calculations are presented in the following subsections. Where available, experimental data has been added for comparison.

\subsection{Bond lengths and dissociation energies}

The vNz basis sets were used to calculate the equilibrium bond lengths ($r_e$) of group 15 nitrides and group 16 carbides, as well as the dissociation energies ($D_e$) of group 16 carbides. Calculations of $D_e$ for group 15 nitrides are not included here, because the $n$p$^3$ atomic states would require a multireference approach, which is outside the scope of this study.

Electrons in the valence shells were included in the correlated calculations, and the virtual space cutoff was set at 100~a.u., which was found to be well above the threshold to obtain converged results. \added{To give an indication of how much of the virtual space is included with this cutoff, Table \ref{Group16C_virtuals_frozen} lists the number of active and frozen virtual orbitals for the group 16 carbides for the basis sets used in the calculations.}

\begin{table}[!htbp]
 \caption{\added{The number of active and frozen virtual orbitals used in the single-point energy calculations at the CCSD(T) level for the group 16 carbides, as a function of basis set.}}\label{Group16C_virtuals_frozen}
 \begin{tabular}{llrrrr}
 \hline
 \noalign{\vspace{4pt}}
 \added{Molecule} & \added{Virtuals}  & \added{v2z} & \added{v3z} & \added{v4z} & \added{v5z} \\
 \noalign{\vspace{4pt}}
 \hline
 \noalign{\vspace{4pt}}
     \added{ CS} &     \added{Active} &  \added{47} & \added{ 80} & \added{130} & \added{205} \\
         &     \added{Frozen} &  \added{16} & \added{ 32} & \added{ 49} & \added{ 62} \\
 \noalign{\vspace{4pt}}
     \added{CSe} &     \added{Active} &  \added{60} & \added{100} & \added{154} & \added{219} \\
         &     \added{Frozen} &  \added{36} & \added{63 } & \added{ 93} & \added{137} \\
 \noalign{\vspace{4pt}}
     \added{CTe} &     \added{Active} &  \added{67} & \added{102} & \added{159} & \added{235}  \\
         &     \added{Frozen} &  \added{58} & \added{ 97} & \added{124} & \added{155}  \\
 \noalign{\vspace{4pt}}
     \added{CPo} &     \added{Active} &  \added{95} & \added{132} & \added{189} & \added{257}  \\
         &     \added{Frozen} &  \added{103} & \added{148} & \added{190} & \added{229}  \\
 \noalign{\vspace{4pt}}
     \added{CLv} &     \added{Active} &  \added{96} & \added{130} & \added{185} & \added{263}  \\
         &     \added{Frozen} &  \added{126} & \added{172} & \added{218} & \added{266}  \\
 \noalign{\vspace{4pt}}
 \hline
 \end{tabular}
\end{table}

The equilibrium bond lengths were determined from the minimum of a 7th order polynomial, obtained from a least-squares fit to calculated single-point energies at internuclear distance intervals of 0.05 \AA\ over an approximately 1~\AA\ range near the equilibrium. We found the equilibrium bond lengths to be converged with respect to the polynomial order, and the number and interval of the points. The dissociation energies were determined by performing a counterpoise calculation on the molecule, with an internuclear distance equal to the calculated equilibrium bond length at the same level of theory and basis set.

\begin{table*}[!htbp]
 \caption{Bond lengths (\AA) of group 15 nitrides as a function of basis set size and correlation level.} \label{Group15N_R_e}
 \begin{tabular}{llccccccl}
 \hline
 \noalign{\vspace{4pt}}
 Molecule & Method  & v2z & v3z & v4z & v5z & CBS(4z) & CBS(5z) & Exp. \\ 
 \noalign{\vspace{4pt}}
 \hline
 \noalign{\vspace{4pt}}
 \noalign{\vspace{4pt}}
      PN &      SCF &   1.4631 &   1.4531 &   1.4484 &   1.4453 &   1.4484 &   1.4453 &  \\
         &     CCSD &   1.5097 &   1.4936 &   1.4858 &   1.4816 &   1.4840 &   1.4807 &  \\
         &  CCSD(T) &   1.5228 &   1.5070 &   1.4990 &   1.4948 &   1.4972 &   1.4938 & 1.4909~\cite{nist-diatomic-webbook} \\
 \noalign{\vspace{4pt}}
     AsN &      SCF &   1.5704 &   1.5685 &   1.5671 &   1.5666 &   1.5671 &   1.5666 &  \\
         &     CCSD &   1.6231 &   1.6171 &   1.6132 &   1.6119 &   1.6116 &   1.6111 &  \\
         &  CCSD(T) &   1.6380 &   1.6329 &   1.6289 &   1.6277 &   1.6274 &   1.6269 & 1.6184~\cite{nist-diatomic-webbook} \\
 \noalign{\vspace{4pt}}
     SbN &      SCF &   1.7775 &   1.7734 &   1.7710 &   1.7694 &   1.7710 &   1.7694 &  \\
         &     CCSD &   1.8476 &   1.8388 &   1.8339 &   1.8315 &   1.8327 &   1.8309 &  \\
         &  CCSD(T) &   1.8714 &   1.8631 &   1.8580 &   1.8556 &   1.8569 &   1.8550 & 1.835~\cite{jenouvrier1978spectre} \\
 \noalign{\vspace{4pt}}
     BiN &      SCF &   2.0021 &   1.9753 &   1.9739 &   1.9725 &   1.9739 &   1.9725 &  \\
         &     CCSD &   1.9769 &   1.9586 &   1.9571 &   1.9556 &   1.9566 &   1.9554 &  \\
         &  CCSD(T) &   2.0024 &   1.9808 &   1.9791 &   1.9777 &   1.9786 &   1.9775 & 1.9349~\cite{cooke2004microwave} \\
 \noalign{\vspace{4pt}}
 \hline
 \end{tabular}
\end{table*}

\begin{figure}[!htbp]
    \centering
    \includegraphics[width=0.85\columnwidth]{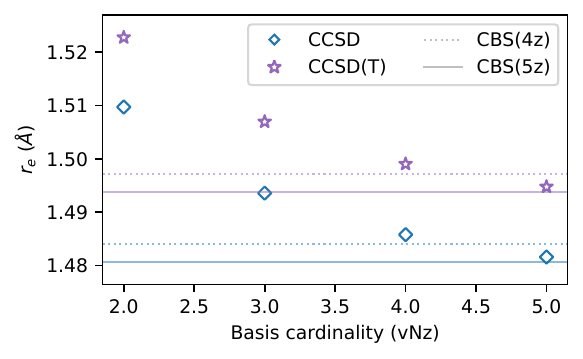}
    \caption{Bond length of PN using the CCSD and CCSD(T) methods as a function of basis set cardinality. Markers indicate the calculated values, and the CBS(4z) and CBS(5z) extrapolations are shown as dotted and solid lines, respectively.}
    \label{fig:BL_PN_CCSD+CCSD(T)_combined}
\end{figure}

The bond lengths \added{$r_e$} of group 15 nitrides are given in Table~\ref{Group15N_R_e}. Focusing on the CCSD(T) results, going from the 4z to the 5z basis set is found to decrease the bond length by 1.3--4.3~m\AA, depending on the molecule (2.4~m\AA\ on average), while the effect on the CBS limit extrapolation is smaller: 0.6--3.4~m\AA\ (1.7~m\AA\ on average).
For PN, these benchmarks are further illustrated in Fig.~\ref{fig:BL_PN_CCSD+CCSD(T)_combined}. A smooth decrease of $r_e$ towards the CBS limit is observed for both the CBS(4z) and CBS(5z) extrapolations. The influence of the additional calculation with the 5z basis set is clear, as it results in a 3.4~m\AA\ correction of $r_e$ at the CBS limit.

The $r_e$ calculations are in good agreement with experimental \added{$r_e$} reference data, with the extrapolated CCSD(T) results within a few m\AA\ from experiment for PN and AsN. For SbN and BiN, however, the difference is larger, at 20~m\AA\ and 43~m\AA\ respectively. The remaining discrepancies with respect to the experimental values can be attributed mostly to higher-order correlation effects and, in particular for the dimers containing a heavy element, higher-order relativistic corrections that are not included in our benchmarks.

The bond lengths and dissociation energies calculated for group 16 carbides can be found in Table~\ref{Group16C_R_e} and Table~\ref{Group16C_D_e}.
\begin{table*}[!htbp]
 \caption{Bond lengths (\AA) of group 16 carbides as a function of basis set and correlation level.} \label{Group16C_R_e}
 \begin{tabular}{llccccccl}
 \hline
 \noalign{\vspace{4pt}}
 Molecule & Method  & v2z & v3z & v4z & v5z & CBS(4z) & CBS(5z) & Exp. \\ 
 \hline
 \noalign{\vspace{4pt}}
      CS &      SCF &   1.5238 &   1.5140 &   1.5109 &   1.5088 &   1.5109 &   1.5088 &  \\
         &     CCSD &   1.5498 &   1.5360 &   1.5306 &   1.5275 &   1.5290 &   1.5267 &  \\
         &  CCSD(T) &   1.5617 &   1.5478 &   1.5423 &   1.5391 &   1.5407 &   1.5383 & 1.534941~\cite{nist-diatomic-webbook} \\
 \noalign{\vspace{4pt}}
     CSe &      SCF &   1.6532 &   1.6498 &   1.6483 &   1.6480 &   1.6483 &   1.6480 &  \\
         &     CCSD &   1.6809 &   1.6747 &   1.6711 &   1.6700 &   1.6696 &   1.6693 &  \\
         &  CCSD(T) &   1.6960 &   1.6898 &   1.6860 &   1.6850 &   1.6845 &   1.6842 & 1.67647~\cite{nist-diatomic-webbook} \\
 \noalign{\vspace{4pt}}
     CTe &      SCF &   1.8790 &   1.8760 &   1.8728 &   1.8719 &   1.8728 &   1.8719 &  \\
         &     CCSD &   1.9079 &   1.9018 &   1.8970 &   1.8954 &   1.8959 &   1.8948 &  \\
         &  CCSD(T) &   1.9270 &   1.9204 &   1.9155 &   1.9139 &   1.9144 &   1.9133 & -- \\
 \noalign{\vspace{4pt}}
     CPo &      SCF &   1.9900 &   1.9813 &   1.9804 &   1.9799 &   1.9804 &   1.9799 &  \\
         &     CCSD &   2.0241 &   2.0104 &   2.0080 &   2.0069 &   2.0070 &   2.0064 &  \\
         &  CCSD(T) &   2.0355 &   2.0219 &   2.0199 &   2.0190 &   2.0193 &   2.0186 & -- \\
 \noalign{\vspace{4pt}}
     CLv &      SCF &   2.2143 &   2.2036 &   2.2016 &   2.2012 &   2.2016 &   2.2012 &  \\
         &     CCSD &   2.2732 &   2.2567 &   2.2525 &   2.2512 &   2.2511 &   2.2505 &  \\
         &  CCSD(T) &   2.3013 &   2.2823 &   2.2774 &   2.2762 &   2.2756 &   2.2753 & -- \\
 \noalign{\vspace{4pt}}
 \hline
 \end{tabular}
\end{table*}

\begin{table*}[!htbp]
 \caption{Dissociation energies (eV) of group 16 carbides as a function of basis set and correlation level.}\label{Group16C_D_e}
 \begin{tabular}{llrrrrrr}
 \hline
 \noalign{\vspace{4pt}}
 Molecule & Method  & v2z & v3z & v4z & v5z & CBS(4z) & CBS(5z) \\
 \noalign{\vspace{4pt}}
 \hline
 \noalign{\vspace{4pt}}
      CS &      SCF &   5.7272 &   5.9932 &   6.0553 &   6.0996 &   6.0553 &   6.0996 \\
         &     CCSD &   6.6326 &   7.0693 &   7.2686 &   7.3648 &   7.3687 &   7.4157 \\
         &  CCSD(T) &   6.8389 &   7.3183 &   7.5269 &   7.6264 &   7.6338 &   7.6807 \\
 \noalign{\vspace{4pt}}
     CSe &      SCF &   4.5162 &   4.6612 &   4.6937 &   4.6983 &   4.6937 &   4.6983 \\
         &     CCSD &   5.4981 &   5.8100 &   5.9686 &   6.0248 &   6.0606 &   6.0711 \\
         &  CCSD(T) &   5.7374 &   6.0968 &   6.2683 &   6.3300 &   6.3698 &   6.3810 \\
 \noalign{\vspace{4pt}}
     CTe &      SCF &   2.9795 &   3.1388 &   3.1825 &   3.1924 &   3.1825 &   3.1924 \\
         &     CCSD &   3.9934 &   4.2811 &   4.4259 &   4.4811 &   4.4997 &   4.5178 \\
         &  CCSD(T) &   4.2931 &   4.6330 &   4.7920 &   4.8529 &   4.8761 &   4.8947 \\
 \noalign{\vspace{4pt}}
     CPo &      SCF &   2.7803 &   2.9735 &   2.9961 &   3.0013 &   2.9961 &   3.0013 \\
         &     CCSD &   2.9068 &   3.1853 &   3.2903 &   3.3326 &   3.3505 &   3.3625 \\
         &  CCSD(T) &   3.1612 &   3.4996 &   3.6230 &   3.6724 &   3.6965 &   3.7090 \\
 \noalign{\vspace{4pt}}
     CLv &      SCF &   4.7664 &   4.9093 &   4.9312 &   4.9376 &   4.9312 &   4.9376 \\
         &     CCSD &   1.8195 &   2.0072 &   2.0806 &   2.1088 &   2.1182 &   2.1277 \\
         &  CCSD(T) &   2.0543 &   2.2965 &   2.3881 &   2.4238 &   2.4389 &   2.4493 \\
 \noalign{\vspace{4pt}}
 \hline
 \end{tabular}
\end{table*}

Similar trends are found as for the group 15 nitrides for both properties. Smooth convergence is obtained when going to the newest basis sets from the 2z, 3z and 4z results. 

\begin{figure}[!htbp]
    \centering
    \includegraphics[width=0.85\columnwidth]{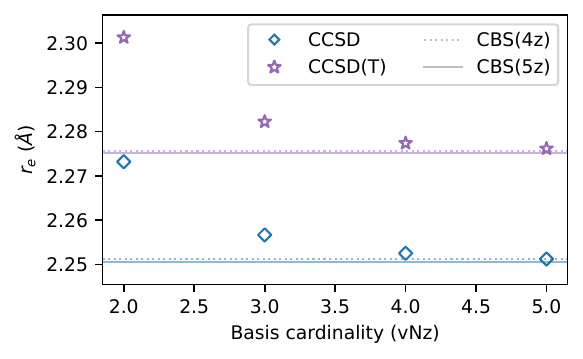}
    \caption{Bond length of CLv using the CCSD and CCSD(T) methods as a function of basis set cardinality. Markers indicate the calculated values, and the CBS(4z) and CBS(5z) extrapolations are shown as dotted and solid lines, respectively.}
    \label{fig:BL_CLv_CCSD+CCSD(T)_combined}
\end{figure}

For the bond lengths, the 5z results and CBS limits contribute 0.3--3.2~m\AA\ and 0.3--2.4~m\AA\ respectively at the CCSD(T) level.
The latter can be considered a rough estimate to the remaining uncertainty with respect to the CBS limit convergence. The calculated bond lengths of CLv at different levels of theory as a function of basis set cardinality are shown in Fig.~\ref{fig:BL_CLv_CCSD+CCSD(T)_combined}, where both well-behaved convergence and very good agreement between the CBS(4z) and CBS(5z) results can be observed.

\begin{figure}[!htbp]
    \centering
    \includegraphics[width=0.85\columnwidth]{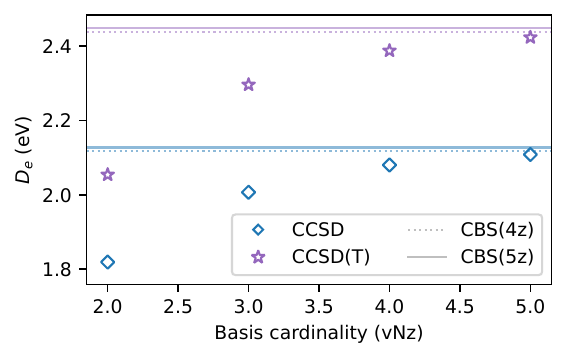}
    \caption{Dissociation energy of CLv using the CCSD and CCSD(T) methods as a function of basis set cardinality. Markers indicate the calculated values, and the CBS(4z) and CBS(5z) extrapolations are shown as dotted and solid lines, respectively.}
    \label{fig:DE_CLv_CCSD+CCSD(T)_combined}
\end{figure}

As expected, the dissociation energy is much more sensitive to the basis set than the bond length. Nevertheless, the dissociation energies show a consistent behavior towards the CBS limit. The 5z basis set increases the dissociation energy by about 1.3\% at the CCSD(T) level, when compared to the 4z result, while the difference between the two CBS limit extrapolations is on average 0.3\%, and up to 50 meV. The large effect on the extrapolated dissociation energies illustrates the importance of using the new 5z basis sets for this property.  The $D_e$ results for CLv are illustrated in Fig.~\ref{fig:DE_CLv_CCSD+CCSD(T)_combined}.

 \added{The effect of the counterpoise correction is illustrated in Table~\ref{CSe_D_e_CP} for the dissociation energy of the CSe molecule. The counterpoise correction is negligible at the SCF level where the basis set is saturated in the s and p spaces, but is about 3~meV at the correlated level when the 5z basis set is used for the CBS extrapolation. This is due to the decreasing degree of saturation of the basis set as the angular momentum  of the basis functions increases. If the difference between the CBS(4z) and CBS(5z) results can be taken as a guide to the error in the CBS(5z) results, the counterpoise correction is smaller than the error by about an order of magnitude. }

\begin{table*}[!htbp]
 \caption{\added{Dissociation energy (eV) of CSe with and without counterpoise (CP) correction, as a function of basis set and correlation level.}}\label{CSe_D_e_CP}
 \begin{tabular}{lrrrrrrr}
 \hline
 \noalign{\vspace{4pt}}
 \added{Method} & \added{CP} & \added{v2z} & \added{v3z} & \added{v4z} & \added{v5z} & \added{CBS(4z)} & \added{CBS(5z)} \\
 \noalign{\vspace{4pt}}
 \hline
 \noalign{\vspace{4pt}}
    \added{SCF} & \added{No} & \added{4.535005} & \added{4.661353} & \added{4.693746} & \added{4.698255} & \added{4.693746} & \added{4.698255} \\ 
     & \added{Yes} & \added{4.516196} & \added{4.661178} & \added{4.693719} & \added{4.698253} & \added{4.693719} & \added{4.698253} \\ 
     & \added{$\Delta_\text{CP}$} & \added{-0.018809} & \added{-0.000175} & \added{-0.000027} & \added{-0.000002} & \added{-0.000027} & \added{-0.000002} \\ 
 \noalign{\vspace{4pt}}
    \added{CCSD} & \added{No} & \added{5.672013} & \added{5.857927} & \added{5.986390} & \added{6.033243} & \added{6.056495} & \added{6.068149} \\ 
     & \added{Yes} & \added{5.498117} & \added{5.810033} & \added{5.968645} & \added{6.024807} & \added{6.060643} & \added{6.071092} \\ 
     & \added{$\Delta_\text{CP}$} & \added{-0.173896} & \added{-0.047894} & \added{-0.017745} & \added{-0.008436} & \added{0.004148} & \added{0.002943} \\ 
 \noalign{\vspace{4pt}}
    \added{CCSD(T)} & \added{No} & \added{5.918655} & \added{6.148209} & \added{6.287356} & \added{6.339025} & \added{6.365257} & \added{6.377800} \\ 
     & \added{Yes} & \added{5.737430} & \added{6.096821} & \added{6.268347} & \added{6.329982} & \added{6.369769} & \added{6.380985} \\ 
     & \added{$\Delta_\text{CP}$} & \added{-0.181225} & \added{-0.051388} & \added{-0.019009} & \added{-0.009043} & \added{0.004512} & \added{0.003185} \\ 
 \noalign{\vspace{4pt}}
 \hline
 \end{tabular}
\end{table*}

For both sets of molecules the bond lengths increase monotonically when going from the light to heavy systems, due to the increasing size of the valence $n$p$_{3/2}$ orbital with $Z$. This is also reflected in the decreasing dissociation energies of the heavier group 16 carbides. 


\subsection{Ionization potentials and electron affinities}
The ionization potentials (IP) and electron affinities (EA) are suitable atomic properties for testing the new basis sets, since they require calculations of the energies of the positive ion and the anion as well as the neutral system. For reliable calculations on these systems, the basis sets must be well-balanced and not biased towards any particular charge state. That said, it is usually necessary to augment a basis set designed for a neutral atom with diffuse functions to represent the more extended charge distribution of an anion; these functions will have a negligible effect on the quality of the calculations for the neutral and the positive systems. 

For the IP calculations, the cvNz basis sets were used for both the neutral and the positive atoms, and electrons in the $n$ and ($n-1$) shells were correlated. The avNz basis sets, which contain the necessary diffuse functions for the description of the anion, were used in the calculations on both the neutral atom and the anion to obtain the EAs; in these calculations only electrons in the $n$ shell were correlated.

\begin{figure}[!htbp]
    \centering
    \includegraphics[width=0.85\columnwidth]{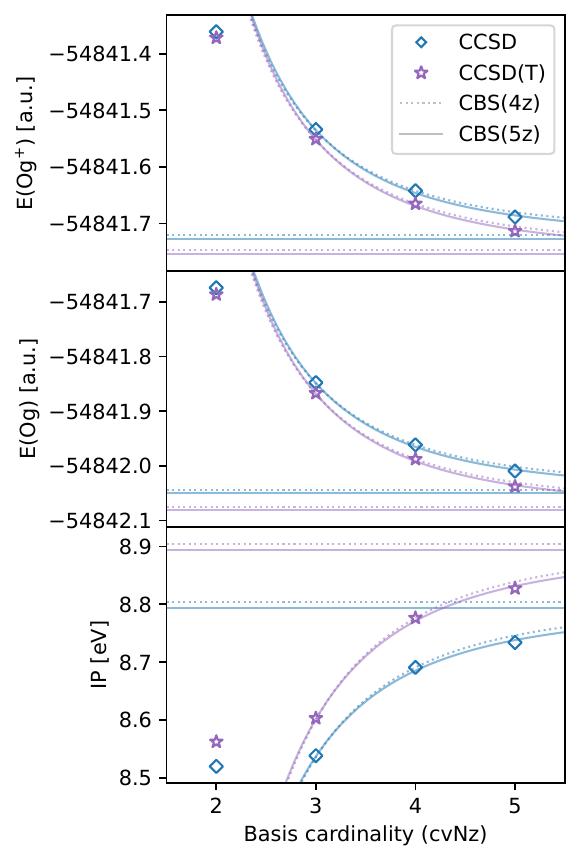}
    \caption{Total energies of Og and its ion (top panels) and IP (bottom panel) as a function of basis set cardinality, calculated with CCSD and CCSD(T). Markers indicate the calculated values, and the CBS(4z) and CBS(5z) extrapolations are shown as dotted and solid lines, respectively.}
    \label{fig:IP+E-E_Og_CCSD+CCSD(T)_combined}
\end{figure}

\begin{figure}[!htbp]
    \centering
    \includegraphics[width=0.85\columnwidth]{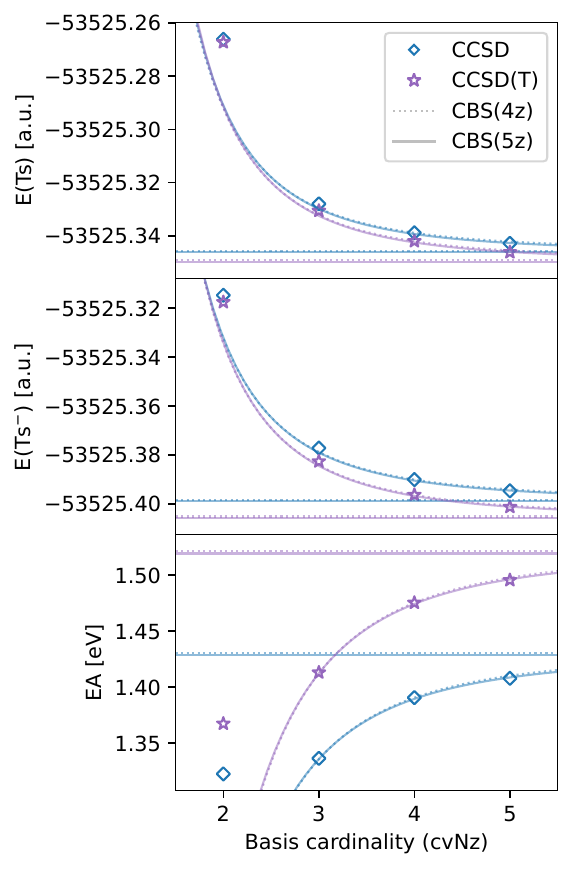}
    \caption{Total energies of Ts and its anion (top panels) and EA (bottom panel) as a function of basis set cardinality, calculated with CCSD and CCSD(T). Markers indicate the calculated values, and the CBS(4z) and CBS(5z) extrapolations are shown as dotted and solid lines, respectively.}
    \label{fig:EA+E-E_Ts_CCSD+CCSD(T)_combined}
\end{figure}

To examine the calculated properties and their CBS limit extrapolation, the IP of Og and the EA of Ts are plotted in Fig.~\ref{fig:IP+E-E_Og_CCSD+CCSD(T)_combined} and Fig.~\ref{fig:EA+E-E_Ts_CCSD+CCSD(T)_combined}. In both cases, the energies of the atom and the ion decrease when using a larger basis set, as expected. In contrast, both the IP of Og and the EA of Ts increase with the basis set size. This is due to the fact that the energy of the neutral (anion) species decreases more sharply than that of the ion (neutral) species for the IP (EA) with increase of the size of the basis set. This, in turn, originates mainly in the larger correlation contribution for the case with the extra electron, which increases with basis set size. We have found that this behavior is consistent in all of our benchmarks.

The results of the IP benchmarks are summarized in Table~\ref{Group18_IPs} for noble gases and in Table~\ref{Group14_IPs} for group 14 elements. Calculated EAs for group 17 elements can be found in Table~\ref{Group17_EAs}.
\begin{table*}[!htbp]
 \caption{IPs (eV) of noble gases as a function of basis set and correlation level. \deleted{The experimental values are taken from the NIST database.~\cite{NIST_ASD}} No experiment is available for Og.}
 \label{Group18_IPs}
 \begin{tabular}{llccccccl}
 \hline
 \noalign{\vspace{4pt}}
 Atom & Method  & cv2z & cv3z & cv4z & cv5z & CBS(4z) & CBS(5z) & Exp.~\added{\cite{NIST_ASD}} \\ 
 \hline
 \noalign{\vspace{4pt}}
He & SCF & 23.4460 & 23.4482 & 23.4482 & 23.4482 & 23.4482 & 23.4482 &  \\
 & CCSD & 24.3807 & 24.5280 & 24.5664 & 24.5792 & 24.5923 & 24.5921 &  \\
 & CCSD(T) & 24.3807 & 24.5280 & 24.5664 & 24.5792 & 24.5923 & 24.5921 & 24.587389011 \\
 \noalign{\vspace{4pt}}
Ne & SCF & 19.8313 & 19.8255 & 19.8250 & 19.8249 & 19.8250 & 19.8249 &  \\
 & CCSD & 21.0278 & 21.2454 & 21.3820 & 21.4361 & 21.4820 & 21.4867 &  \\
 & CCSD(T) & 21.0663 & 21.3017 & 21.4504 & 21.5102 & 21.5592 & 21.5651 & 21.564541 \\
 \noalign{\vspace{4pt}}
Ar & SCF & 14.7809 & 14.7567 & 14.7548 & 14.7547 & 14.7548 & 14.7547 &  \\
 & CCSD & 15.1840 & 15.4344 & 15.5890 & 15.6373 & 15.7033 & 15.6966 &  \\
 & CCSD(T) & 15.1910 & 15.4793 & 15.6480 & 15.7015 & 15.7726 & 15.7661 & 15.7596119 \\
 \noalign{\vspace{4pt}}
Kr & SCF & 13.2660 & 13.2402 & 13.2385 & 13.2384 & 13.2385 & 13.2384 &  \\
 & CCSD & 13.5332 & 13.6990 & 13.8357 & 13.8878 & 13.9368 & 13.9392 &  \\
 & CCSD(T) & 13.5431 & 13.7466 & 13.8942 & 13.9512 & 14.0031 & 14.0065 & 13.9996055 \\
 \noalign{\vspace{4pt}}
Xe & SCF & 11.7029 & 11.6741 & 11.6731 & 11.6731 & 11.6731 & 11.6731 &  \\
 & CCSD & 11.7033 & 11.8238 & 11.9695 & 12.0227 & 12.0765 & 12.0774 &  \\
 & CCSD(T) & 11.7181 & 11.8698 & 12.0273 & 12.0865 & 12.1430 & 12.1454 & 12.1298437 \\
 \noalign{\vspace{4pt}}
Rn & SCF & 11.1752 & 11.1520 & 11.1512 & 11.1512 & 11.1512 & 11.1512 &  \\
 & CCSD & 10.3572 & 10.4731 & 10.5877 & 10.6372 & 10.6719 & 10.6793 &  \\
 & CCSD(T) & 10.3817 & 10.5267 & 10.6530 & 10.7082 & 10.7457 & 10.7545 & 10.74850 \\
 \noalign{\vspace{4pt}}
Og & SCF & 11.5579 & 11.5324 & 11.5300 & 11.5299 & 11.5300 & 11.5299 &  \\
 & CCSD & 8.5199 & 8.5383 & 8.6909 & 8.7339 & 8.8040 & 8.7933 &  \\
 & CCSD(T) & 8.5624 & 8.6031 & 8.7763 & 8.8275 & 8.9045 & 8.8945 & -- \\
 \noalign{\vspace{4pt}}
 \hline
 \end{tabular}
\end{table*}
The calculated IPs for the noble gases change considerably between 4z and 5z, about 50~meV on average when looking at the CCSD(T) values, although this corresponds to a less drastic relative effect of 0.4\%, since the IPs are quite high for noble gases. 
Fortunately, the CBS(4z) and CBS(5z) are in good agreement with each other and with the experimental values, illustrating the reliability of the extrapolation scheme.
\begin{table*}[]
    \caption{IPs (eV) of group 14 elements as a function of basis set and correlation level. \deleted{The experimental values are taken from Ref.~\onlinecite{CRC}. }No experiment is available for Fl.}
    \label{Group14_IPs}
 \begin{tabular}{llccccccl}
 \hline
 \noalign{\vspace{4pt}}
 Atom & Method  & cv2z & cv3z & cv4z & cv5z & CBS(4z) & CBS(5z) & Exp.~\added{\cite{CRC}} \\ 
 \hline
 \noalign{\vspace{4pt}}
C & SCF & 9.9971 & 9.9954 & 9.9951 & 9.9951 & 9.9951 & 9.9951 &  \\
 & CCSD & 10.7534 & 10.8905 & 10.9159 & 10.9237 & 10.9347 & 10.9334 &  \\
 & CCSD(T) & 10.8926 & 11.0713 & 11.1083 & 11.1204 & 11.1355 & 11.1344 & 11.26030 \\
 \noalign{\vspace{4pt}}
Si & SCF & 7.1095 & 7.1044 & 7.1043 & 7.1042 & 7.1043 & 7.1042 &  \\
 & CCSD & 7.7124 & 7.8274 & 7.8472 & 7.8529 & 7.8617 & 7.8605 &  \\
 & CCSD(T) & 7.8688 & 8.0094 & 8.0369 & 8.0449 & 8.0570 & 8.0554 & \phantom{0}8.15169 \\
 \noalign{\vspace{4pt}}
Ge & SCF & 6.8830 & 6.8757 & 6.8755 & 6.8755 & 6.8755 & 6.8755 &  \\
 & CCSD & 7.4815 & 7.5611 & 7.5796 & 7.5890 & 7.5932 & 7.5956 &  \\
 & CCSD(T) & 7.6664 & 7.7534 & 7.7748 & 7.7863 & 7.7907 & 7.7940 & \phantom{0}7.8994 \\
 \noalign{\vspace{4pt}}
Sn & SCF & 6.3770 & 6.3676 & 6.3674 & 6.3674 & 6.3674 & 6.3674 &  \\
 & CCSD & 6.9937 & 7.0648 & 7.0900 & 7.1027 & 7.1085 & 7.1117 &  \\
 & CCSD(T) & 7.1379 & 7.2129 & 7.2427 & 7.2578 & 7.2646 & 7.2685 & \phantom{0}7.3439 \\
 \noalign{\vspace{4pt}}
Pb & SCF & 6.1698 & 6.1607 & 6.1606 & 6.1604 & 6.1606 & 6.1604 &  \\
 & CCSD & 7.1503 & 7.2475 & 7.2767 & 7.2897 & 7.2981 & 7.3003 &  \\
 & CCSD(T) & 7.2320 & 7.3354 & 7.3713 & 7.3869 & 7.3976 & 7.4000 & \phantom{0}7.41666 \\
 \noalign{\vspace{4pt}}
Fl & SCF & 6.2514 & 6.2563 & 6.2558 & 6.2558 & 6.2558 & 6.2558 &  \\
 & CCSD & 8.3987 & 8.4741 & 8.6298 & 8.6560 & 8.7438 & 8.7180 &  \\
 & CCSD(T) & 8.4236 & 8.4938 & 8.6628 & 8.6936 & 8.7865 & 8.7606 & \phantom{0}-- \\
    \noalign{\vspace{4pt}}
    \hline
    \end{tabular}
\end{table*}

In case of the group 14 elements, the same comparison between the 4z and the 5z values yields a difference of about 16 meV on average, corresponding to 0.2\%,  and  an even  smaller difference between the CBS(4z) and CBS(5z) results. The remaining difference from experiment of 0.1~eV  can be attributed to missing higher order correlation and relativistic effects. 

\begin{table*}[]
    \caption{EAs (eV) of group 17 elements as a function of basis set and correlation level.}\label{Group17_EAs}
 \begin{tabular}{llrrrrrrl}
 \hline
 \noalign{\vspace{4pt}}
 Molecule & Method  & av2z & av3z & av4z & av5z & CBS(4z) & CBS(5z) & Exp. \\ 
 \noalign{\vspace{4pt}}
 \hline
 \noalign{\vspace{4pt}}
F & SCF & 1.3402 & 1.3337 & 1.3331 & 1.3330 & 1.3331 & 1.3330 &  \\
 & CCSD & 3.0577 & 3.1190 & 3.1790 & 3.1989 & 3.2232 & 3.2217 &  \\
 & CCSD(T) & 3.1958 & 3.2835 & 3.3519 & 3.3747 & 3.4021 & 3.4007 & 3.4011898(24) \cite{F_Br_EA} \\
 \noalign{\vspace{4pt}}
Cl & SCF & 2.5493 & 2.5258 & 2.5244 & 2.5243 & 2.5244 & 2.5243 &  \\
 & CCSD & 3.3339 & 3.3673 & 3.4636 & 3.4884 & 3.5350 & 3.5261 &  \\
 & CCSD(T) & 3.3864 & 3.4579 & 3.5610 & 3.5895 & 3.6372 & 3.6296 & 3.612725(28) \cite{Cl_EA} \\
 \noalign{\vspace{4pt}}
Br & SCF & 2.4051 & 2.3758 & 2.3747 & 2.3747 & 2.3747 & 2.3747 &  \\
 & CCSD & 3.0867 & 3.1192 & 3.2194 & 3.2458 & 3.2934 & 3.2848 &  \\
 & CCSD(T) & 3.1318 & 3.1974 & 3.3030 & 3.3328 & 3.3808 & 3.3737 & 3.363588(3) \cite{F_Br_EA} \\
 \noalign{\vspace{4pt}}
I & SCF & 2.2032 & 2.1853 & 2.1847 & 2.1846 & 2.1847 & 2.1846 &  \\
 & CCSD & 2.7814 & 2.8139 & 2.9217 & 2.9497 & 3.0007 & 2.9915 &  \\
 & CCSD(T) & 2.8168 & 2.8831 & 2.9965 & 3.0286 & 3.0796 & 3.0722 & 3.0590465(37) \cite{I_EA} \\
 \noalign{\vspace{4pt}}
At & SCF & 1.5394 & 1.5219 & 1.5214 & 1.5214 & 1.5214 & 1.5214 &  \\
 & CCSD & 2.1517 & 2.1699 & 2.2663 & 2.2923 & 2.3370 & 2.3296 &  \\
 & CCSD(T) & 2.1885 & 2.2405 & 2.3431 & 2.3728 & 2.4182 & 2.4121 & 2.41578(7) \cite{LeiKarGuo20} \\
 \noalign{\vspace{4pt}}
Ts & SCF & 3.5496 & 3.5293 & 3.5286 & 3.5284 & 3.5286 & 3.5284 &  \\
 & CCSD & 1.3226 & 1.3365 & 1.3907 & 1.4080 & 1.4308 & 1.4288 &  \\
 & CCSD(T) & 1.3674 & 1.4131 & 1.4754 & 1.4957 & 1.5214 & 1.5195 & -- \\
    \noalign{\vspace{4pt}}
    \hline
    \end{tabular}
\end{table*}
The benefit of the 5z basis set is clearly visible for the calculated EAs, as the difference compared to the 4z calculations is about 20--30~meV in each case. That said, the agreement between the two CBS values is very good, with a difference of only 5~meV on average, or 0.2\%. Nevertheless, 5~meV is a significant contribution in the context of spectroscopy, where single meV accuracy is the desired objective.

The results presented here show the importance of the newly developed basis sets for increasing the accuracy and reliability of correlated calculations. 

\section{Conclusions}
Relativistic quintuple-zeta basis sets for the p-block elements have been presented. The basis sets include SCF-optimized functions for the occupied spinors, while valence and core correlating functions were optimized in multireference SDCI calculations on the ground configuration. Additionally, diffuse functions optimized on the anion, or derived from neighboring elements for group 18, are also provided.

The new basis sets were tested by calculating the equilibrium bond lengths and dissociation energies of diatomic molecules and ionization potentials and electron affinities of atoms. Smooth convergence to the basis set limit is observed with increased basis set quality from the previously available double-zeta, triple-zeta, and quadruple-zeta basis sets from the same family to the newly developed quintuple-zeta basis sets. Comparison of CBS limit schemes shows that the addition of these basis sets has a significant effect on the accuracy, which can be on the order of several~m\AA\ for bond lengths, and several~meV for ionization potentials and electron affinities, while the effect on the dissociation energy is more significant, on the order of tens of meV. Use of these basis sets in combination with state-of-the-art approaches for treatment of relativity and correlation will allow significantly increased accuracy in calculations on heavy elements and their compounds.

\deleted{The basis sets will be made available on zenodo.org when this manuscript is published.}

\begin{acknowledgments}
Dutch Research Council (NWO) project number Vi.Vidi.192.088 and LISA: European Union’s H2020 Framework Programme under grant agreement no. 861198. We thank the Center for Information Technology of the University of Groningen for their support and for providing access to the Hábrók high performance computing cluster.
\end{acknowledgments}

\section*{Author declarations}

\subsection*{Conflict of Interest}
The authors have no conflicts to disclose.

\subsection*{Author Contributions}
\noindent
\textbf{Marten L. Reitsma}: Investigation; Methodology; Visualization; Writing – original draft; Writing – review \& editing.
\textbf{Eifion H. Prinsen}: Investigation; Writing – original draft; Writing – review \& editing.
\textbf{Johan D. Polet}: Investigation; Writing – review \& editing.
\textbf{Anastasia Borschevsky}: Funding acquisition; Supervision; Methodology; Writing – original draft; Writing – review \& editing.
\textbf{Kenneth G. Dyall}: Investigation; Methodology; Software; Supervision; Writing – original draft; Writing – review \& editing.

\subsection*{DATA AVAILABILITY}
\added{The basis sets are available on zenodo.org at the following URL (\href{https://doi.org/10.5281/zenodo.18351835}{doi.org/10.5281/zenodo.18351835}), and will also be available on the Basis Set Exchange (https://www.basissetexchange.org/).}

The data that support the findings of this study are available from the corresponding author upon reasonable request.

\section*{References}
\bibliography{cc_basis_refs,dyall_basis_refs,dyall_refs,expt_p5z,more_refs}

\end{document}